\documentclass[conference,10pt]{IEEEtran}
\IEEEoverridecommandlockouts

\usepackage[top=0.75in, bottom=1in, left=0.625in, right=0.625in]{geometry}
\usepackage{breakurl}
\usepackage{cite}
\usepackage{amsmath,amssymb,amsfonts}
\usepackage{algorithmic}
\usepackage{graphicx}
\usepackage{textcomp}
\usepackage{color}
\usepackage[lofdepth,lotdepth]{subfig}
\usepackage{soul}
\usepackage{float}
\usepackage{optidef}
\usepackage{multirow}
\usepackage{hhline}

\begin{document}
%
\title{Developing an NTN Architecture for End-to-End Performance Evaluation.}
%
%
%
\author{Md Mahfuzur Rahman~\IEEEmembership{Senior Member,~IEEE}, Nishith Tripathi ~\IEEEmembership{Senior Member,~IEEE},\\ Jeffrey H. Reed~\IEEEmembership{Fellow,~IEEE} , and Lingjia Liu ~\IEEEmembership{ Fellow,~IEEE}
\thanks{M. M. Rahman, N. Tripathi, J. H. Reed, and L. Liu are with Wireless@Virginia Tech, Bradley Department of Electrical and Computer Engineering, Virginia Tech, VA, USA. (Corresponding author's e-mail: mrahman2@vt.edu). This research is sponsored in part by the NSF project titled "Distributed space and terrestrial networking infrastructure for multi-constellation coexistence" (NSF CCRI award number 2235139). This research is supported in part by the NTIA PWSCIF project titled "A holistic cybersecurity testing framework project" (NTIA PWSCIF award number: 51-60-IF007). This material is also based upon work supported by the National Science Foundation under grant CNS-2148212 and is supported in part by funds from federal agency and industry partners as specified in the Resilient \& Intelligent NextG Systems (RINGS) program.}}
%
%
%

\maketitle
\begin{abstract}
Non-Terrestrial Networks (NTN) are emerging as critical enablers of global connectivity, particularly in remote, unserved, underserved, or maritime regions lacking traditional infrastructure. While much of the existing work on NTN focuses on theoretical or simulated evaluations, practical implementations remain limited. In this paper, we present SpaceNET, a transparent NTN testbed that leverages the Starlink Low Earth Orbit (LEO) satellite constellation in conjunction with Mininet-based emulation to perform end-to-end performance assessments across real-world maritime and terrestrial endpoints that can also be applied to 5th generation (5G). Specifically, we establish a bidirectional link between a ground terminal located in Blacksburg, Virginia, and a maritime terminal aboard a cruise ship near Key West, Florida. We report detailed transmission control protocol (TCP) throughput, user datagram protocol (UDP) throughput, and latency measurements using two different user terminals - a) Smartphone, and b) very small aperture terminal (VSAT), emphasizing the transparent nature of the NTN payload, where the satellite acts solely as a relay node. Our results provide new insights into the performance limits and reliability of commercial LEO-based NTN applications. The SpaceNET testbed offers a reproducible and extensible platform for future research in NTN routing, mobility support, and cross-layer optimization.

\end{abstract}
\vspace{-0.4cm}
\section{Introduction}
Non-terrestrial networks (NTN) will serve as a cornerstone of future 5G-Advanced and 6G systems by providing high-quality, omnipresent services to both unserved and underserved regions. NTN refer to wireless communication systems that leverage aerial and spaceborne platforms, spanning from Low Altitude Platforms (LAPs) and High Altitude Platforms (HAPs) to Low Earth Orbit (LEO) and High Earth Orbit (HEO) satellites \cite{mahfuz_tcom, Ju-Hyung}. These networks are anticipated to play a transformative role in both civilian and military domains, enhancing navigation, global communication, and remote sensing capabilities. Looking ahead to the 6G era, NTN is envisioned to support three fundamental use cases: (i) delivering ubiquitous connectivity to rural, airborne, and maritime users, (ii) ensuring service continuity by extending terrestrial network (TN) coverage, (iii) enabling scalable broadcast and multicast services to large user populations, and insuring security for military communications \cite{milcom, Nasir, Zoheb_NTN_3}. Recent standardization efforts, such as 3GPP TR 38.863 Release 17, have focused on the coexistence and interoperability of NTN and TN, particularly within the sub-6 GHz bands \cite{3gpp863}. Complementary earlier releases, including 3GPP TR 38.821 and TR 38.811, have laid essential groundwork for NTN-TN operations across both mmWave and sub-6 GHz frequencies \cite{3gpp821, 3gpp811}.

The increasing momentum around NTN deployments is underscored by the major investments of industry leaders such as SpaceX, Amazon, OneWeb, Hughes, and Iridium. For example, SpaceX’s Starlink program has already launched over 8,000 LEO satellites, with plans to scale to more than 12,000—and potentially up to 46,400—to provide global broadband coverage \cite{spacex}. Amazon’s Project Kuiper, aiming to launch more than 3,000 LEO satellites, has partnered with NTT DOCOMO, NTT Communications, and SKY Perfect JSAT to deliver advanced satellite connectivity services in Japan \cite{amazon, amazon2}. Hughes, a dominant force in the Americas, continues to enhance its GEO satellite capacity, with the Jupiter series recently reaching up to 500 Gbps \cite{hughes}.

The work in \cite{lin} analyzes and evaluates the throughput and capacity performance of LEO NTN systems using 5G New Radio (NR), highlighting that while individual satellites can deliver downlink capacities of approximately 600Mbps in the S band (30MHz) for handheld terminals and 7Gbps in the Ka band (400MHz) for VSATs. However, their study is limited within the theoretical realm. The authors in \cite{imstn} evaluated the balance among energy efficiency, data rates, and spectral utilization when integrating satellites into terrestrial 6G networks. In \cite{willey_ntn}, LEO satellite constellations were applied for traffic offloading and backhauling delay-tolerant IoT data, while \cite{uav2ntn} explored the offloading of UAV communications to NTN, yielding a significant reduction in both UAV downlink outages and terrestrial uplink failures. Further, \cite{orbital_edge} proposed an orbital edge computing paradigm using LEO satellite constellations to offload computing tasks, addressing latency challenges for both real-time and non-real-time workloads.

Given the inherent heterogeneity between TN and NTN systems, efficient resource management is essential to unlock their full potential in 6G networks. For example, in space-air-ground integrated networks (SAGIN), the uneven distribution of communication and computing resources hinders the delivery of consistent quality-of-service (QoS), especially for latency- or outage-sensitive applications \cite{bodong_sag}. Various studies have explored critical integration challenges, including the need for carefully designed timing alignment in satellite-based 5G NR random access \cite{Shuxun}, the adaptation of random access periodicity for NTN NB-IoT due to its large coverage footprint \cite{carla}, and strategies to mitigate co-channel interference in NTN-TN shared-spectrum environments \cite{Lee}. Notably, experimental platforms, such as the one developed in \cite{yifei} using SDRs and embedded systems, have begun to reveal the performance intricacies and optimization opportunities of real NTN deployments.

Building on this body of work, we present SpaceNET, an experimental transparent NTN testbed leveraging the commercial Starlink LEO constellation and software-defined network emulation to evaluate end-to-end performance between a terrestrial gNB and a maritime UE. By establishing a live NTN link between a ground terminal in Blacksburg, Virginia, and a cruise ship terminal near Key West, Florida, SpaceNET enables reproducible experiments assessing transmission control protocol (TCP), user datagram protocol (UDP), and Internet Control Message Protocol (ICMP) behavior under real-world satellite-ground conditions. Our results offer critical empirical insights into the performance dynamics of transparent NTN systems and inform the development of robust, scalable NTN-augmented 5G and 6G networks.

.

\section{Motivation and Contribution}
\subsection{Motivations}
The demand for ubiquitous, high-capacity wireless connectivity has never been greater, driven by the rapid proliferation of data-hungry applications, the expansion of Internet of Things (IoT) deployments, and the growing needs of mobility-centric use cases such as maritime communications, aviation, and remote terrestrial operations. While terrestrial cellular networks, including 4G Long Term Evolution (LTE) and 5G NR, have made significant advances in spectral efficiency and coverage, they remain inherently limited by ground-based infrastructure, leaving vast oceanic, rural, and unserved/underserved regions without reliable broadband access.

NTN, particularly those leveraging LEO satellite constellations, have emerged as a promising solution to bridge these connectivity gaps. LEO satellites offer lower latency, higher throughput, and greater coverage flexibility compared to legacy geostationary systems, making them suitable for delay-sensitive applications and dynamic user environments. However, despite the theoretical promise, there exists a critical lack of real-world experimental testbeds capable of evaluating the performance of NTN systems under practical operating conditions, especially for end-to-end communication scenarios spanning maritime and terrestrial domains.

This gap is further accentuated by the predominance of transparent satellite architectures, where the satellite functions solely as a relay node without performing onboard processing. While such designs reduce satellite complexity and cost, they place increased demands on the terrestrial segment for routing, adaptation, and optimization. Understanding the cross-layer performance interactions between physical-layer impairments, transport protocols, and routing strategies in transparent NTN is therefore essential for designing robust, scalable, and efficient satellite-integrated networks.

Motivated by these challenges, we developed SpaceNET, a novel transparent NTN testbed leveraging commercial Starlink LEO infrastructure and software-defined emulation tools. This platform enables the first-of-its-kind real-world performance evaluation of maritime-to-ground NTN links, providing actionable insights into protocol behavior, bottleneck dynamics, and optimization opportunities across system layers.
\vspace*{0.3cm}

\subsection{Contributions}
In this work, we present SpaceNET, a transparent NTN testbed that bridges terrestrial and maritime domains using the commercial Starlink LEO satellite network and a Mininet-based emulation platform. Our setup enables direct experimentation with real TCP/UDP IP traffic and allows for fine-grained control over routing, logging, and link-state visualization. This work provides critical insights into practical deployment challenges and achievable performance in transparent NTN links, offering a foundation for future NTN protocol development and cross-layer optimization. The specific contributions of this paper are summarized as follows.

\begin{enumerate}

    \item We design and implement SpaceNET, a practical transparent NTN testbed combining Starlink and Mininet to evaluate real-world performance.
    
    \item We conduct live maritime-to-ground experiments measuring TCP/UDP throughput and ICMP latency across a bidirectional NTN link.  

    \item We demonstrate the feasibility of transparent satellite payload evaluation using commercial off-the-shelf (COTS) infrastructure without onboard satellite processing.

    \item We provide an open framework to support further studies in NTN routing, scheduling, and application-aware optimization.
    
 \end{enumerate}

\section{System Model}

\subsection{System Overview}

\begin{figure}[h]
\centering
\includegraphics[width=3in, height=2in]{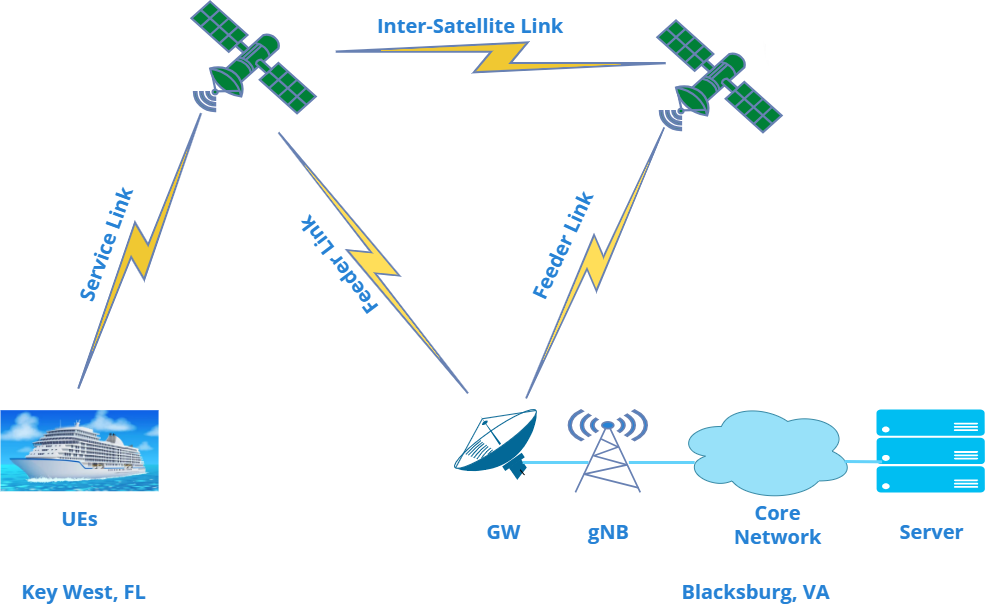} 
\caption{A realistic NTN network (Transparent payload)}
\label{system-model}
\end{figure}

This study investigates an integrated cellular-NTN architecture that combines a terrestrial network (TN) next-generation NodeB (gNB) with a commercial LEO satellite constellation, specifically leveraging the Starlink LEO constellation. The primary objective is to deliver enhanced bandwidth and reduced latency services to maritime UE located onboard a cruise ship, as depicted in Figure \ref{system-model}. The NTN operates under a transparent payload architecture, where LEO satellites act solely as relay nodes, forwarding radio signals without performing onboard processing. At the terrestrial segment, the gNB transmits at a power level of 36 dBm, delivering signals to the ground station, which handles up/down-conversion for the feeder link. This gNB is connected to the 5G System (5GS) core network through a high-speed fiber optic backhaul, ensuring seamless integration between NTN and terrestrial domains. Both the service link (UE–satellite) and feeder link (gNB–satellite) adhere to the standardized 3GPP protocol stack, ensuring efficient interoperability and robust communication. Additionally, the inter-satellite links (ISLs) operate in Frequency Range 2 (FR2) within the millimeter-wave (mmWave) band, whereas the gNB/UE–satellite links utilize the K-band, optimizing spectral efficiency across network segments.

The experimental testbed, referred to as SpaceNET, is composed of two principal endpoints connected via the Starlink LEO satellite system. Ground terminal stationed at Blacksburg, Virginia integrates a TN gNB. It also hosts a Mininet emulation environment, enabling the simulation of advanced routing topologies, multi-hop paths, and congestion scenarios. Maritime terminal onboard on a cruise ship near Key West, Florida connects directly with the Starlink satellite. We establish a bidirectional end-to-end link between these terminals and employ standardized network performance measurement tools, including iperf3 and ping, to evaluate TCP, UDP, and ICMP performance under realistic NTN conditions. The entire emulated platform is built using Ubuntu LTS 24.04 in a virtual machine. Importantly, the system emulates a transparent NTN architecture, where satellite nodes serve solely as forwarding relays, while all protocol processing is handled at the terrestrial segment. Through configurable IP-layer forwarding rules and static routing in Mininet, we create controlled experimental conditions to assess multi-hop behavior, dynamic routing, and congestion impacts within the NTN framework.
\setlength{\textfloatsep}{0pt}
\begin{table}[h]
\centering
\caption{Different commercial grade testing tools}
\begin{tabular}{|l|c|c|c|}
\hline
Sl.  & Measures & Protocol & Testing tools  \\  
\hline
1 & Latency (RTT), Packet loss & ICMP &  Ping\\
\hline
2 & Bandwidth (Throughput) & TCP & iPerf PDL \\
\hline
3 & Bandwidth (Throughput) & TCP & iPerf PUL  \\
\hline
4 & Bandwidth (Throughput) & UDP &  iPerf PDL \\
\hline
5 & Bandwidth (Throughput) & UDP & iPerf PUL \\
\hline
\end{tabular}
\label{tools}
\end{table}

\subsection{NTN Path Loss Models}

As recommended by the  3GPP TR 38.821 \cite{3gpp821} and 3GPP TR 38.811 \cite{3gpp811}  standards, the NTN channel is inherently different from TN channel, encompassing the integration of path losses caused by the adverse weather conditions such as rain, cloud, fog, and scintillation. For the NTN link, the total path loss \( P_{\mathrm{NTN}} \) (in dB) is given by:
\begin{equation} 
\label{eqNTN_new}
P_{\mathrm{NTN}} = Q_{\mathrm{fspl}} + Q_{\mathrm{entry}} + Q_{\mathrm{atm}} + Q_{\mathrm{scint}},
\end{equation}
where \( Q_{\mathrm{fspl}} \) is the line-of-sight (LOS) free space path loss, \( Q_{\mathrm{entry}} \) is the building entry loss, \( Q_{\mathrm{atm}} \) accounts for gaseous attenuation, and \( Q_{\mathrm{scint}} \) is the scintillation loss. According to 3GPP TR 38.811, gaseous losses mainly affect frequencies above 52\,GHz; since the focus here is on the K-band (below 52\,GHz), \( Q_{\mathrm{atm}} \) can be neglected. Assuming clear-sky, outdoor conditions, \( Q_{\mathrm{entry}} \) and \( Q_{\mathrm{scint}} \) are also disregarded. For an elevation angle of $70^0$, we have used a shadowing effect of 2.6dB in our simulation. Then the basic NTN path loss becomes:
\begin{equation} \label{eqBasic_new}
Q_{\mathrm{fspl}} = 32.45 + 20 \log_{10}(r) + 20 \log_{10}(f),
\end{equation}
where \( f \) is the frequency in GHz and \( r \) is the distance in meters.

\section{Performance Evaluation}
\subsection{Simulation Setup}
We present a comprehensive cellular system architecture that seamlessly integrates TN base stations (BSs) with satellite constellations, enabling flexible interoperability across multiple Radio Access Technologies (RATs), including 4G LTE, 5G NR, and beyond. While the system design accommodates diverse RAT configurations, the scope of our analysis is confined to E2E NTN link-level simulations. In our system model, the satellite is configured as a transparent payload; however, with appropriate adaptations, the onboard gNB can be reconfigured to operate as a regenerative payload, offering additional processing capabilities. Consistent with the specifications in 3GPP TR 38.901 \cite{3gpp901}, TN gNBs are assumed to transmit at a power level of 36dBm at 2GHz, after which the signal undergoes frequency up/down conversion for transmission over the feeder link across extended distances. A complete summary of all simulation parameters is provided in Table \ref{parameters}.
%
%
%
%
%
%
\setlength{\textfloatsep}{0pt}
\begin{table}
\centering
\caption{Simulation parameters.}
\begin{tabular}{|l|c|c|}
\hline
Sl. & Features & Parameters \\  
\hline
1 & Constellation & Starlink LEO \\
\hline
2 & Elevation angle & $70^0$ \\
\hline
3 & Channel freq ISLs & 37.0 GHz \\
\hline
4 & Channel freq sat to ground & 12.7 GHz \\
\hline
5 & Channel freq ground to sat & 14.5 GHz \\
\hline
6 & Channel bandwidth downlink & 240 MHz \\
\hline
7 & Channel bandwidth uplink & 60 MHz \\
\hline
8 & Polarization loss & 3 dBi \\
\hline
9 & Misalignment attenuation loss & 0.5 dB \\
\hline
10 & Starlink merit figure & 9.2 dB/K \\
\hline
11 & Satellite EIRP & 80.9 dBm \\
\hline
12 & Satellite EIRP (dbW) & 50.9 dBW \\
\hline
13 & Base station tx power & 36.0 dBm \\
\hline
14 & Ground station rx antenna gain & 33.2 dBi \\
\hline
15 & Ground station tx antenna gain & 34.6 dBi \\
\hline
16 & UE types & Smartphone \& VSAT \\
\hline
17 & UE tx antenna gain & 0 \& 43.2 dBi \\
\hline
18 & UE rx antenna gain & 0 \& 39.7 dBi \\
\hline
19 & UE tx power & 23 \& 33 dBm \\
\hline
20 & NTN payload type & Transparent payload \\
\hline
\end{tabular}
\label{parameters}
\end{table}
%
%
%
%
%

\subsection{Performance Analysis}

\begin{figure}[h]
    \centering
    \subfloat[\centering Ping terminal output (First run)]
    {\includegraphics[width=3in, height=1.5in]{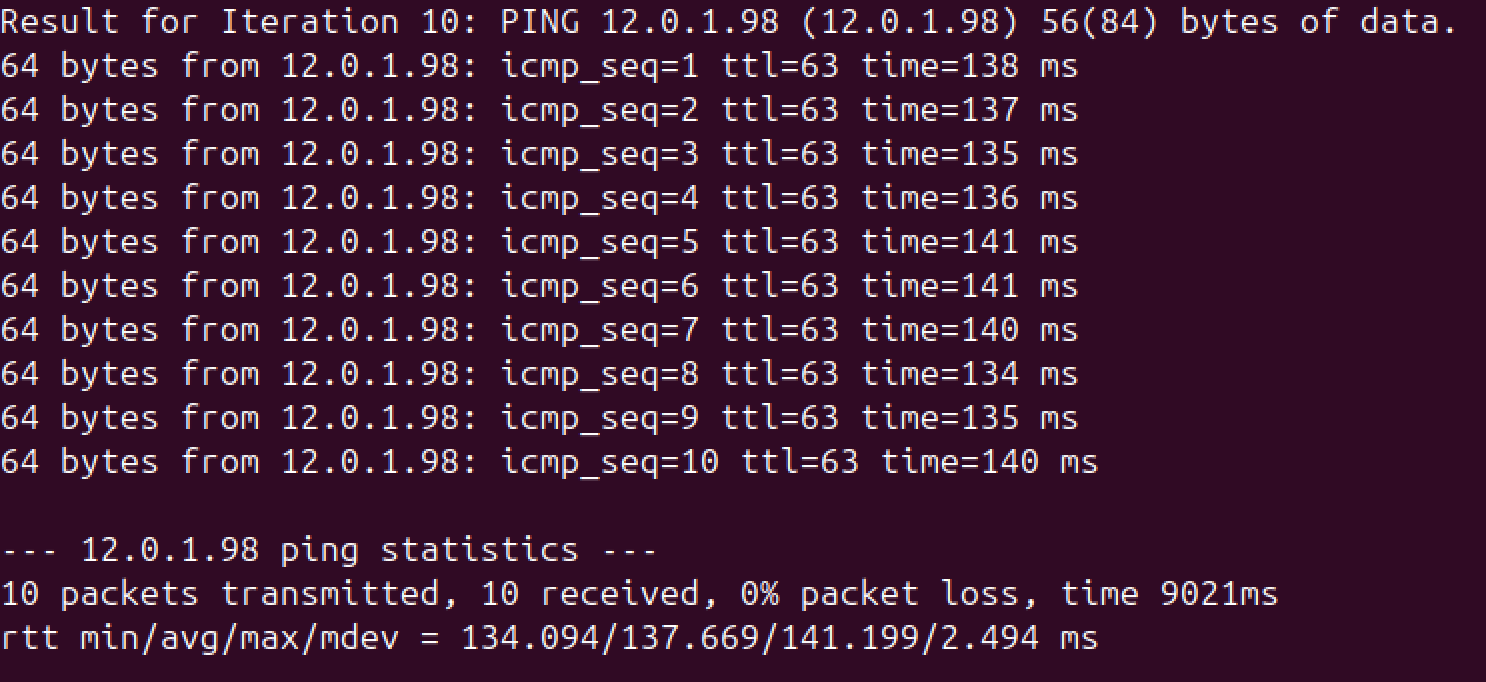}\label{ping_test}}%
    \vspace{2pt}
     \subfloat[\centering Ping plot (Second run)]
    {\includegraphics[width=3in, height=1.5in]{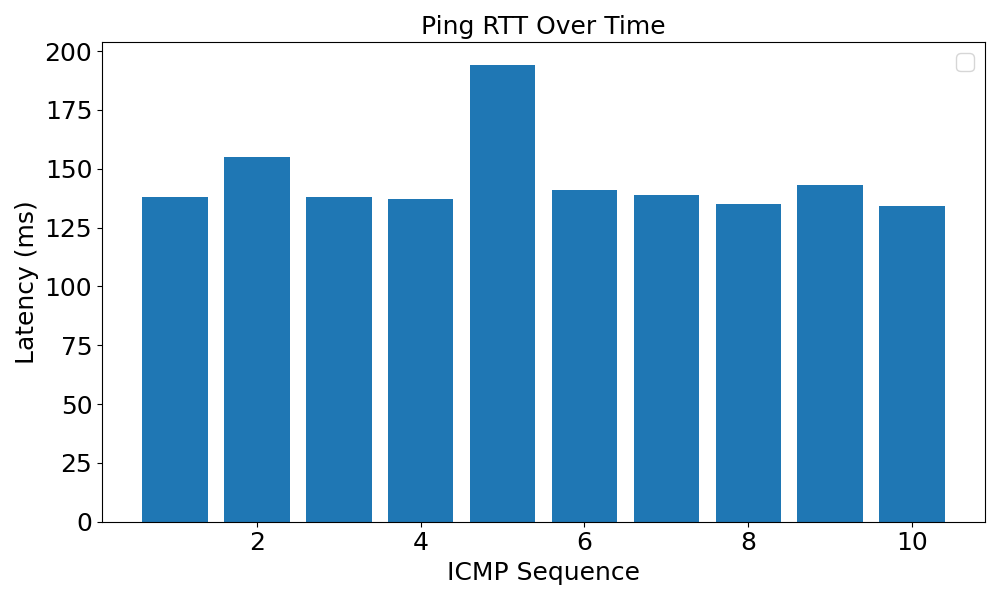}\label{ping_bar}}%
    \caption{\centering End-to-end ping testing.}
    \label{ping}
\end{figure}

The testbed consisted of a UE, with the downlink and uplink paths routed through a virtualized network and a gNB located in our research lab in Blacksburg, VA. The ping utility was used to generate ICMP echo requests from the UE to the gNB IP address over a satellite-ground backhaul emulator, with ten consecutive ICMP packets transmitted at one-second intervals. The RTT values were logged to measure temporal latency variations and derive key performance metrics.
In this section, we provide simulation results on various metrics for latency and bandwidth aware scenarios.

\subsubsection{End-to-End RTT (Latency) Evaluation}
In this section, we present the round-trip time (RTT) performance of a real-world, long-distance end-to-end (E2E) test between a UE located at Key West, Florida, and a base station (gNB) deployed in Blacksburg, Virginia. Although the current implementation does not incorporate full 5G capabilities, the setup represents a transparent, early-stage NTN scenario, utilizing E2E connectivity to assess RTT behavior across a geographically disjoint wireless link. As LEO satellites provide coverage to certain geographical area for only $\sim$ 7 seconds \cite{3gpp811,3gpp821}, our simulation was limited to only 10 seconds. Fig. \ref{ping_bar} illustrates the measured RTT across the 10 ICMP sequences. The latency values range from a minimum of 134 ms to a maximum of 194 ms, yielding a mean RTT of 145.4 ms and a standard deviation of 18.07 ms. While the observed latencies are higher than those expected in terrestrial deployments (typically $<$ 100 ms), they remain within acceptable bounds for NTN-enabled broadband connectivity. 
\subsubsection{Physical Downlink Throughput}
To rigorously assess the performance characteristics of our transparent NTN architecture, we conducted comparative experiments using TCP and UDP traffic flows over the Starlink LEO satellite link. These experiments measured physical downlink (PDL) throughput between terrestrial and maritime terminals, highlighting how transport-layer protocols interact with satellite link characteristics.

\begin{figure}[h]
    \centering
    \subfloat[\centering iPerf TCP PDL throughput]
    {\includegraphics[width=3in, height=1.5in]{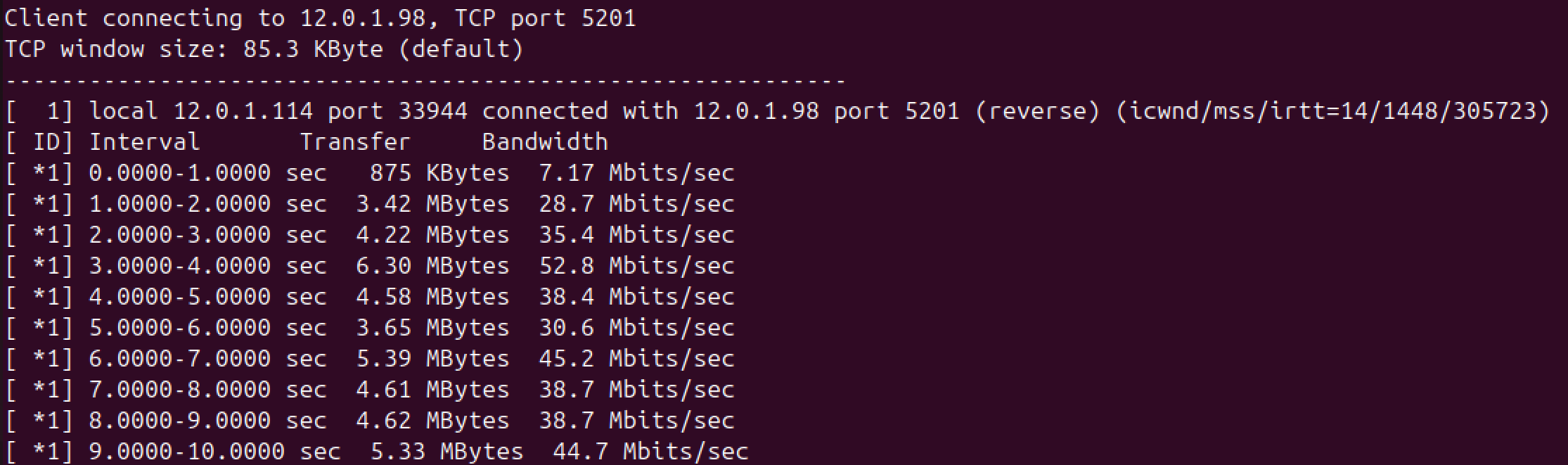}\label{tcp}}%
    \vspace{2pt}
     \subfloat[\centering iPerf UDP PDL throughput]
    {\includegraphics[width=3in, height=1.5in]{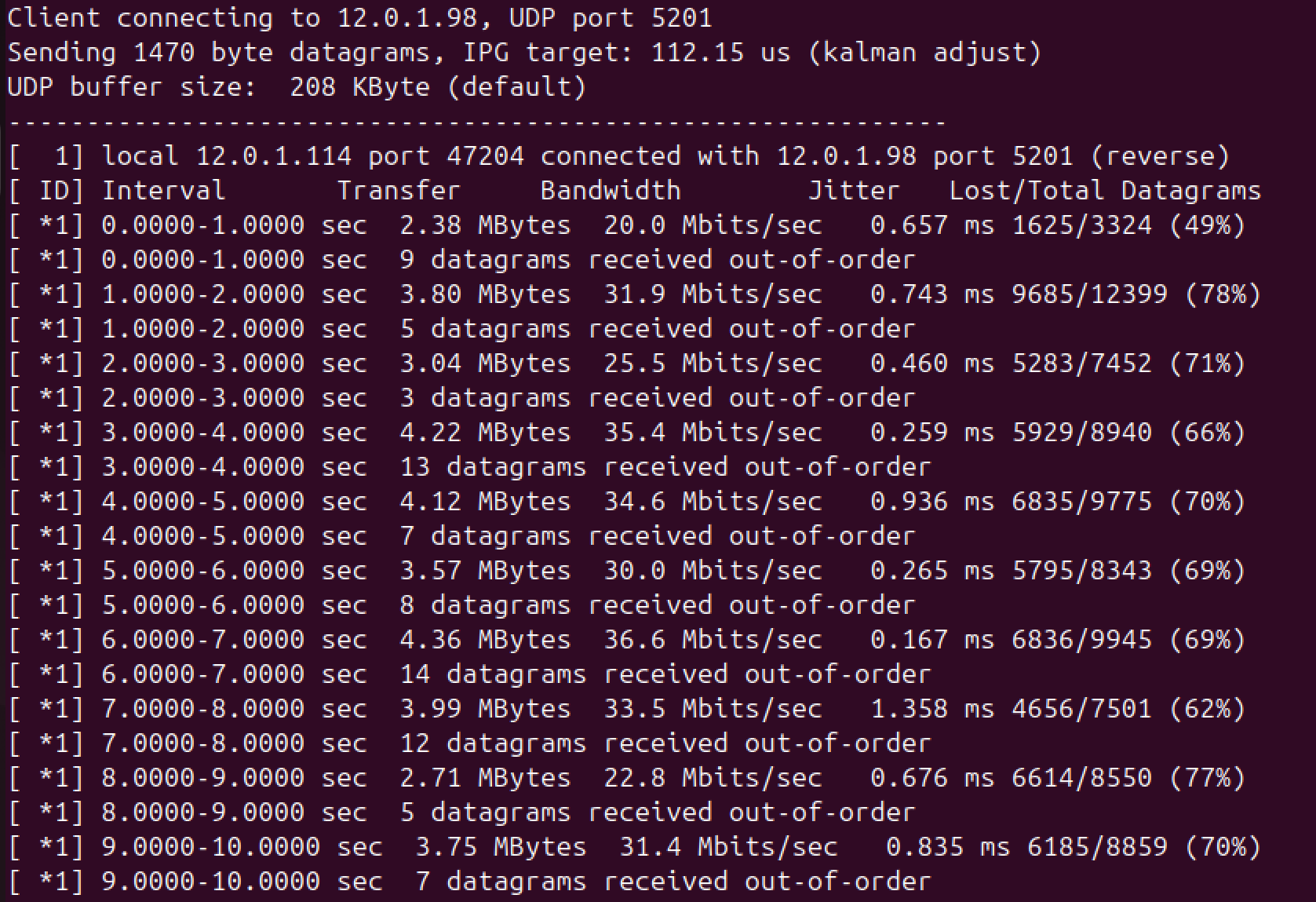}\label{udp}}%
    \caption{\centering End-to-end iPerf PDL testing.}
    \label{pdl}
\end{figure}

As shown in Fig. \ref{tcp} TCP measurements revealed moderate but variable throughput over time, reflecting the well-known sensitivity of TCP to link conditions. The highest instantaneous TCP throughput reached 52.8 Mbps, with most intervals fluctuating between 35 Mbps and 45 Mbps, and observed drops to as low as 7.14 Mbps. These results suggest that while TCP remains the default choice for reliable data transfer, its inherent design struggles under high-latency NTN links. By contrast, as depicted in Fig. \ref{udp} UDP flows consistently above 30 Mbps over the test duration, which is bit opposite, could be associated with lower bandwidth utilization for UDP. Our experiments reinforce the conclusion that UDP-based protocols, when paired with appropriate application-layer error handling (e.g., forward error correction or custom reliability schemes), are well suited for transparent NTN links where maximizing raw data delivery is prioritized.

\subsubsection{Physical Uplink Throughput}
\begin{figure}[h]
    \centering
    \subfloat[\centering iPerf TCP PUL Throughput]
    {\includegraphics[width=3in, height=1.5in]{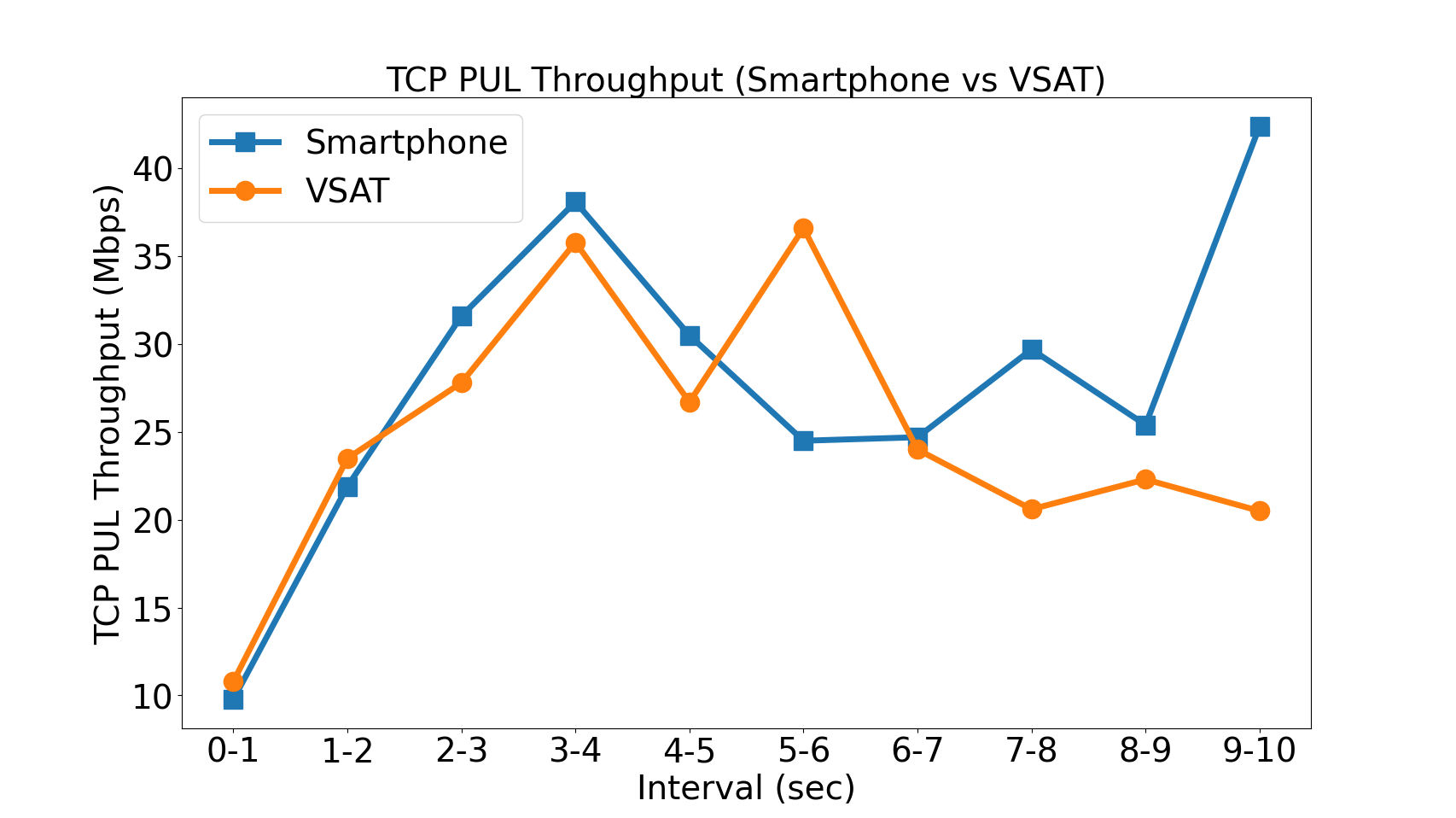}\label{tcp_pul}}%
    \vspace{2pt}
     \subfloat[\centering iPerf UDP PUL Throughput]
    {\includegraphics[width=3in, height=1.5in]{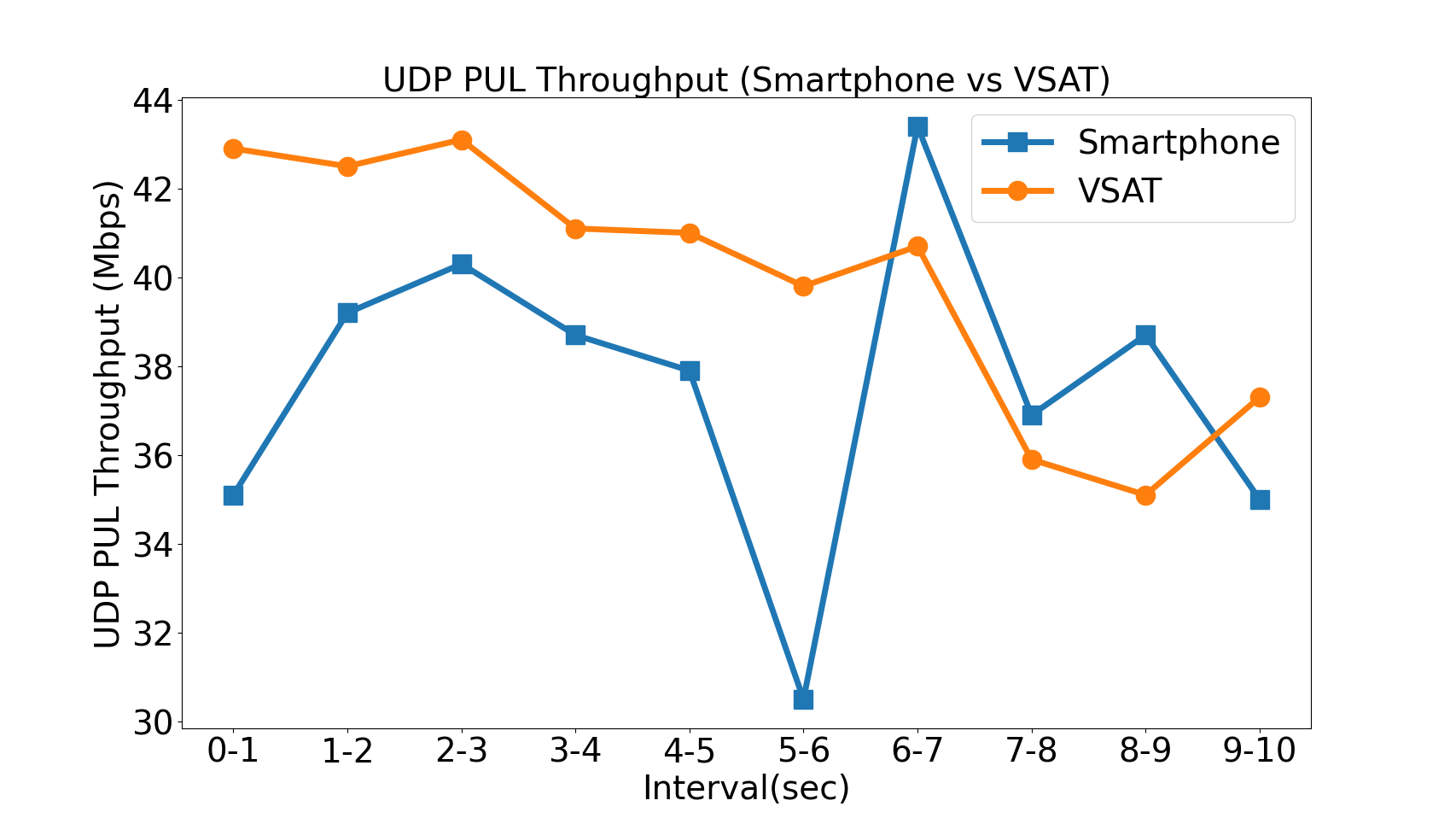}\label{udp_pul}}%
    \caption{\centering End-to-end iPerf PUL testing.}
    \label{pul}
\end{figure}

We conducted a detailed comparative study of physical uplink (PUL) performance between two terminal types — smartphone-based terrestrial UE and Very Small Aperture Terminal (VSAT) — under both TCP and UDP traffic profiles. Fig. \ref{tcp_pul} and \ref{udp_pul} summarize the measured throughput across 10-second intervals, highlighting protocol-dependent differences and the impact of terminal characteristics. As shown in Fig. \ref{tcp_pul}, the TCP uplink throughput for the smartphone consistently outperformed the VSAT link, despite the smartphone’s inherently lower transmit power and antenna gain. The smartphone achieved a peak TCP throughput of   $\sim$42 Mbps, compared to $\sim$36 Mbps for the VSAT. 

These findings highlight the importance of cross-layer interactions, particularly the impact of physical-layer characteristics on transport-layer throughput. In contrast, Fig. \ref{udp_pul} shows that VSAT outperformed the smartphone under UDP uplink conditions, achieving near-saturated and stable throughput around 40–43 Mbps across most intervals. The smartphone, while competitive, displayed noticeable dips—particularly a sharp drop at interval 5–6 seconds to $\sim$30 Mbps—suggesting transient loss events or application-layer throttling. These differences stem from the absence of congestion control in UDP, where link-layer stability and buffer availability dominate performance issues. 
\section{Planned Future Work}
UE transmit power in LTE is controlled through two key offsets: the \textit{global power offset} ($P_{0,\text{global}}$) and the \textit{individual power offset} ($P_{0,\text{individual}}$), both set during the Radio Resource Control (RRC) configuration stage. The base station adjusts these offsets over time as network conditions evolve, informed by measurement reports such as channel quality indicator (CQI) and signal-to-noise ratio (SNR). Effective power control is pivotal for minimizing inter-user interference, a critical limiting factor in achieving high spectral efficiency in dense network deployments.

The theoretical spectral efficiency attainable by a user under these conditions is described by \cite{mahfuz_tcom}:
\begin{equation}
\Gamma_{m,n}(b_k) = \log_2\left(1 + \frac{z_{m,n}^{b_k} \cdot \theta_{m,n}^{b_k}}{\sum_{\substack{n' = 1, n' \neq n}}^{N} \gamma_{m,n}^{b_k} z_{m,n'}^{b_k} \theta_{m,n}^{b_k} + \delta^2} \right),
\end{equation}
where $z_{m,n}^{b_k}$ denotes the transmit power and $\theta_{m,n}^{b_k}$ the channel gain between the $m$-th user and $n$-th base station for resource block group $b_k$, $\gamma_{m,n}^{b_k}$ is the binary association indicator, and $\delta^2$ represents the background noise power. This expression illustrates the pronounced role of aggregate interference on achievable data rates.

The system’s transmit power budget is bounded by \cite{mahfuz_tcom}:
\begin{equation}
\text (C1) \hspace{10pt} \sum_{b_k=1}^{B_k} \sum_{m=1}^{\mathcal{M}}\gamma_{m,n}^{b_k} z_{m,n}^{b_k} \leq G_{n,\text{max}}, \quad  \forall n \in \mathcal{N},
\end{equation}
where $G_{n,\text{max}}$ specifies the maximum transmit power of the $n$-th base station.

The overall system performance objective is formalized as:
\begin{equation}
\Phi(\mathbf{z}, \mathbf{\gamma}) = \Gamma_{\text{sum}},
\end{equation}
with $\mathbf{z}$ and $\mathbf{\gamma}$ denoting the complete transmit power and association matrices, respectively. The corresponding optimization problem is framed as:
\begin{equation}
\text{P1:} \quad \max_{\mathbf{z} \geq 0, \mathbf{\gamma} \in {0,1}} \Phi(\mathbf{z}, \mathbf{\gamma}) \quad \text{subject to (C1)}.
\end{equation}

Since the problem P1 is computationally intractable (NP-hard), we introduce an alternative fractional programming (FP)-driven algorithm specifically designed for interference mitigation. Our proposed scheme was validated and compared against a state-of-the-art FP baseline~\cite{Wei_Yu}. As demonstrated in~\cite{mahfuz_tcom, milcom}, our method matches the baseline’s performance while offering enhanced structural properties, such as improved convergence behavior and more flexible parameter tuning.

This work also investigates the relaying of RF signals from Blacksburg, VA, to a cruise ship positioned near Key West, FL, where a concentrated user base is anticipated to seek access to cellular services. Due to the substantial number of users within the compact confines of the cruise ship, deploying robust interference mitigation techniques is critical to achieving efficient radio resource management and sustaining reliable connectivity.

Future research will focus on advancing this Mininet-based system by integrating ground-to-satellite links using a Keysight NTN channel emulator, PROPSIM \cite{keysight_propsim_f64}, which will enable precise emulation of realistic RF conditions (including doppler, delay, and multipath) between ground stations and satellite nodes. The testbed will also incorporate a 5G gNB located in Blacksburg, VA, enabling realistic evaluation of routing between satellite nodes and terrestrial endpoints. In this enhanced architecture, Mininet will continue to orchestrate the routing and network topology, while the Keysight emulator will ensure accurate channel modeling across ground-to-space links. Ultimately, the goal is to develop a scalable, interference-resilient power control modeling solution suitable for next-generation 5G and beyond.

\section{Conclusion}
This paper presented SpaceNET, a real-world transparent NTN testbed leveraging the Starlink LEO satellite constellation and Mininet-based emulation to evaluate end-to-end performance across maritime and terrestrial terminals. Through extensive experiments, we analyzed TCP and UDP performance over satellite links, examining protocol behavior, transport-layer limitations, and the influence of physical-layer characteristics. Our findings reveal that a) TCP flows over NTN links exhibit considerable throughput variability, influenced by satellite-induced delay, congestion control dynamics, and transient link conditions, b) UDP flows achieve consistently high throughput, highlighting the absence of built-in congestion responsiveness but raising fairness and reliability concerns in shared environments, c) Smartphone-based UEs, despite lower transmit power and antenna gain compared to VSAT terminals, demonstrated superior TCP uplink performance due to faster congestion window scaling and reduced buffer-induced penalties, and d) VSAT terminals excelled under UDP uplink conditions, benefiting from stable high-gain satellite links and dedicated bandwidth. These results underscore the complex cross-layer interactions that shape NTN system performance and emphasize the need for adaptive, NTN-aware transport protocols to fully exploit the capabilities of modern satellite networks. The SpaceNET framework offers a reproducible, extensible platform for future studies on advanced routing, application-aware optimization, and the integration of emerging NTN standards into terrestrial network ecosystems.

 \bibliography{ntn_opt2}

\end{document}